\documentclass[preprint2,numberedappendix]{emulateapj-rtx4}
\usepackage{graphicx}
\usepackage{times}
\usepackage{url}
\usepackage{bm}
\usepackage{natbib}
\graphicspath{{./fig/}{./png/}}

\newcommand{\EQ}{\begin{equation}}
\newcommand{\EE}{\end{equation}}
\newcommand{\EQA}{\begin{eqnarray}}
\newcommand{\EEA}{\end{eqnarray}}
\newcommand{\brac}[1]{\langle #1 \rangle}
\newcommand{\pd}{\partial}

\newcommand{\mean}[1]{\overline{#1}}

\newcommand{\nut}{\nu_{\rm t}}

\newcommand{\etatz}{\eta_{\rm t0}}

\newcommand{\urms}{u_{\rm rms}}

\newcommand{\brms}{B_{\rm rms}}
\newcommand{\Beq}{B_{\rm eq}}

\newcommand{\kef}{k_{\rm f}}
\newcommand{\kfo}{k_{\rm f}^{(\omega)}}

\newcommand{\chit}{\chi_{\rm t}}
\newcommand{\chitm}{\chi_{\rm tm}}

\newcommand{\Pm}{{\rm Pm}}
\newcommand{\Rm}{{\rm Rm}}
\newcommand{\Rey}{{\rm Re}}

\newcommand{\Pra}{{\rm Pr}}
\newcommand{\Ta}{{\rm Ta}}

\newcommand{\Rat}{{\rm Ra}_{\rm t}}

\newcommand{\Co}{{\rm Co}}

\def\onethird{{\textstyle{1\over3}}}
\def\onehalf{{\textstyle{1\over2}}}

\begin{document}

\title{Cyclic magnetic activity due to turbulent convection in spherical wedge geometry}

\author{Petri J.\ K\"apyl\"a$^{1}$, Maarit J.\ Mantere$^{1}$ and Axel Brandenburg$^{2,3}$}
\affil{$^1$Physics Department, Gustaf H\"allstr\"omin katu 2a, PO Box 64,
FI-00014 University of Helsinki, Finland\\
$^2$NORDITA, Royal Institute of Technology and Stockholm University,
Roslagstullsbacken 23, SE-10691 Stockholm, Sweden\\
$^3$Department of Astronomy, AlbaNova University Center,
Stockholm University, SE 10691 Stockholm, Sweden}
\email{petri.kapyla@helsinki.fi
($ $Revision: 1.139 $ $)
}

\begin{abstract} 
  We report on simulations of turbulent, rotating, stratified,
  magnetohydrodynamic convection in spherical wedge geometry. An
  initially small-scale, random, weak-amplitude magnetic field is
  amplified by several orders of magnitude in the course of the
  simulation to form oscillatory large-scale fields in the saturated
  state of the dynamo. The differential rotation is solar-like (fast
  equator), but neither coherent meridional poleward circulation nor 
  near-surface shear layer develop in these runs.
  In addition to a poleward branch of magnetic activity beyond 50 degrees
  latitude, we find for the first time a pronounced equatorward branch
  at around 20 degrees latitude, reminiscent of the solar cycle.
\end{abstract}

\keywords{Magnetohydrodynamics -- convection -- turbulence}


\section{Introduction}

The solar magnetic field exhibits a quasi-periodic cycle with a period
of approximately 22 years. This cycle is manifested by the appearance
of sunspots in low latitude activity belts that migrate towards the
equator as the sunspot cycle progresses. Reproducing this behavior
remains a major challenge to theoreticians. Mean-field models, where
small-scale turbulent effects are parameterized \citep[e.g.][]{KR80},
have reproduced many aspects of the solar cycle, but with broadly varying
assumptions for the various parameterizations \citep[see,
e.g.][]{DC99,O03,KKT06,KO11}.

Another, computationally much more demanding, but physically more
consistent route is to solve the equations of magnetohydrodynamics
directly without resorting to ill-defined parameterizations for the
small scales. In practise, however, realistic Reynolds and Rayleigh
numbers, describing the effects of molecular diffusion with respect to
advection, are not accessible to simulations
\citep[e.g.][]{CS86,MT09,K11}. The usual approach is to enhance the
diffusion coefficients to levels that are computationally feasible
while striving to maximize the resolution.

Early spherical shell simulations were able to produce a solar-like
rotation profile, i.e.\ one with ``equatorward acceleration,''
and oscillatory large-scale dynamos
\citep{G83,G85}. However, the direction of propagation of the
dynamo wave was towards the poles, in contradiction to the Sun. This
can be qualitatively explained by a Parker dynamo wave with
positive radial shear near the equator in conjunction
with negative kinetic helicity density, or a positive $\alpha$-effect, in the
northern hemisphere \citep{P55}. More sophisticated
simulations with solar rotation rate and luminosity have failed to
produce strong large-scale magnetic fields \citep{BMT04} or clear
cyclic behavior \citep{Mea11}. These runs omitted a
stable layer below the convection zone. When such a layer is added,
non-oscillatory large-scale fields are found
also for solar parameters \citep{Browning06}.
Later, oscillatory solutions have been obtained from similar
simulations with subgrid-scale modeling \citep{Ghizaru10,Racine11}.
These are the most solar-like
solutions so far, but also in them the activity is at too high
latitudes and the activity belts do not propagate towards the
equator. 
When the rotation rate is increased from the solar value in runs
without an overshoot layer, first stable wreaths of strong large-scale
fields appear \citep{BBBMT10}, and at even more rapid rotation,
poleward migrating activity is found \citep{KKBMT10,BMBBT11}.

We report here results from simulations of turbulent convection in
spherical wedge geometry with solar-like equatorward acceleration that
exhibit, for the first time, equatorward migrating magnetic activity
near the equator and a polar branch at high latitudes.
The numerical model is the same as that in \cite{KMB11} but here
we add magnetic fields.

\section{The model}
\label{sec:model}
We model a segment of a star, i.e.\ a
``wedge'', in spherical polar coordinates, where $(r,\theta,\phi)$
denote radius, colatitude, and longitude. The radial, latitudinal,
and longitudinal extents of the wedge are
$0.7R \leq r \leq R$, $\theta_0 \leq \theta \leq \pi-\theta_0$, and $0
\leq \phi \leq \phi_0$, respectively, where $R$ is the radius of the
star. Here we take $\theta_0=\pi/8$ and $\phi_0=\pi/2$.

We solve the compressible hydromagnetics equations,
\begin{equation}
\frac{\pd \bm A}{\pd t} = {\bm u}\times{\bm B} - \eta \mu_0 {\bm J},
\end{equation}
\begin{equation}
\frac{D \ln \rho}{Dt} = -\bm\nabla\cdot\bm{u},
\end{equation}
\begin{equation}
\frac{D\bm{u}}{Dt} = \bm{g} -2\bm\Omega\times\bm{u}+\frac{1}{\rho}
\left(\bm{J}\times\bm{B}-\bm\nabla p
+\bm\nabla \cdot 2\nu\rho\bm{\mathsf{S}}\right),
\end{equation}
\begin{equation}
T\frac{D s}{Dt} = \frac{1}{\rho}\left[\bm\nabla \cdot (K \bm\nabla T
+ \chit \rho T \bm\nabla s) + 2\nu \bm{\mathsf{S}}^2\right],
\label{equ:ss}
\end{equation}
where ${\bm A}$ is the magnetic vector potential, $\bm{u}$ is the
velocity, ${\bm B} =\bm\nabla\times{\bm A}$ is the magnetic field,
${\bm J} =\mu_0^{-1}\bm\nabla\times{\bm B}$ is the current density,
$\eta$ is the magnetic diffusivity, $\mu_0$ is the vacuum
permeability, $D/Dt = \pd/\pd t + \bm{u} \cdot \bm\nabla$ is the
advective time derivative,
$\nu$ is the kinematic viscosity, $K$ is the radiative heat conductivity,
$\chi_{\rm t}$ is the unresolved turbulent heat conductivity.
$\rho$ is the density, $s$
is the specific entropy, $T$ is the temperature, and $p$ is the
pressure. The fluid obeys the ideal gas law with $p=(\gamma-1)\rho e$,
where $\gamma=c_{\rm P}/c_{\rm V}=5/3$ is the ratio of specific heats
at constant pressure and volume, respectively, and $e=c_{\rm V} T$ is
the internal energy.
The gravitational acceleration is $\bm{g}=-GM\hat{\bm{r}}/r^2$,
where $G$ is the gravitational constant, $M$ is the mass of the star,
and $\hat{\bm{r}}$ is the unit vector in the radial direction. We omit
the centrifugal force \cite[cf.][]{KMGBC11}. The rate of strain tensor
$\bm{\mathsf{S}}$ is given by
$\mathsf{S}_{ij}=\onehalf(u_{i;j}+u_{j;i})
-\onethird \delta_{ij}\bm\nabla\cdot\bm{u}$,
where the semicolons denote covariant differentiation \citep{MTBM09}.

\subsection{Initial and boundary conditions}
\label{sec:initcond}
The initial state is isentropic and the hydrostatic
temperature gradient is $\pd T/\pd r=-g/[c_{\rm V}(\gamma-1)(m+1)]$,
where $m=1.5$ is the polytropic index.
We fix the value of $\pd T/\pd r$ on the lower boundary.
The density profile follows from hydrostatic
equilibrium. The heat conduction profile is chosen so that radiative
diffusion is responsible for supplying the energy flux in the system,
with $K$ decreasing more than two orders of magnitude from bottom to
top \citep{KMB11}.
A weak random small-scale seed magnetic field is taken as initial
condition (see below).

The radial and latitudinal boundaries are taken to be impenetrable and
stress free; see Equations (14) and (15) of \cite{KMGBC11}.
For the magnetic field we assume perfect conductors at the lower
radial and latitudinal boundaries, and radial field at the outer
radial boundary; see Equations (15)--(17) of \cite{KKBMT10}.
On the latitudinal boundaries we assume that the thermodynamic
quantities have zero first derivatives, thus suppressing heat fluxes
through the boundaries.

On the upper boundary we apply a black body condition
\begin{equation}
\sigma T^4  = -K\frac{\pd T}{\pd r} - \chi_{\rm t} \rho T \frac{\pd s}{\pd r},
\end{equation}
where $\sigma$ is the Stefan--Boltzmann constant.
We use a modified value for $\sigma$ that takes into account that our
Reynolds and Rayleigh numbers are much smaller than in reality,
so $K$ and therefore the flux are much larger than in the Sun.

\subsection{Dimensionless parameters}
We obtain non-dimensional quantities by choosing
$R = GM = \rho_0 = c_{\rm P} = \mu_0 = 1$,
where $\rho_0$ is the initial density at $0.7R$.
Our simulations are defined by the energy flux imposed at the bottom
boundary,
$F_{\rm b}=-(K \pd T/\pd r)|_{r=0.7R}$,
the temperature at the top boundary, $T_1=T(r=R)$,
as well as the values of $\Omega_0$, $\nu$, $\eta$, and
$\chitm=\chit(r_{\rm m}=0.85R)$.
The corresponding nondimensional input parameters are the luminosity parameter
$\mathcal{L} = L_0/[\rho_0 (GM)^{3/2} R^{1/2}]$,
the normalized pressure scale height at the surface,
$\xi = [(\gamma-1) c_{\rm V}T_1]R/GM$,
the Taylor number $\Ta=(2\Omega R^2/\nu)^2$,
the Prandtl number $\Pra=\nu/\chitm$,
the magnetic Prandtl number $\Pm=\nu/\eta$,
and the non-dimensional viscosity $\tilde{\nu}=\nu/\sqrt{GMR}$.
Other useful diagnostic parameters are the Reynolds number
$\Rey=\urms/\nu \kef$ and the Coriolis number $\Co=2\Omega_0/\urms \kef$,
where $\urms=\sqrt{(3/2)\brac{u_r^2+u_\theta^2}}$ is the rms velocity.
Note that for $\urms$ we omit the contribution from the azimuthal velocity,
because its value is dominated by effects from the differential rotation
\citep{KMGBC11}.
The Taylor number can also be written as $\Ta=\Co^2\Rey^2(\kef R)^4$,
with $\kef R\approx21$.
Due to the fact that the initial stratification is isentropic, we
quote the (semi-) turbulent Rayleigh number $\Rat$ from the thermally
relaxed state of the run,
\begin{eqnarray}
\Rat\!=\!\frac{GM(\Delta r)^4}{\nu \chitm R^2} \bigg(-\frac{1}{c_{\rm P}}\frac{{\rm d}s}{{\rm d}r} \bigg)_{r_{\rm m}},
\label{equ:Co}
\end{eqnarray}
where $\kef=2\pi/\Delta r$
is an estimate of the wavenumber of the largest eddies,
and $\Delta r=0.3R$ is the thickness of the layer.
The magnetic field is expressed in equipartition field strengths,
$\Beq(r)=\langle \mu_0 \rho \bm{u}^2 \rangle^{1/2}_{\theta\phi}$, where
the subscripts indicate averaging over $\theta$ and $\phi$ 
with azimuthally averaged mean flows subtracted.

The simulations were performed with the {\sc Pencil
  Code}\footnote{http://code.google.com/p/pencil-code/}, which is a
high-order finite difference method for solving the compressible
equations of magnetohydrodynamics.

\section{Results}
\label{sec:results}
Our primary simulation (Run~B4m) is continued from a thermally relaxed
snapshot of a hydrodynamic Run~B4 of \cite{KMB11} with
$\mathcal{L}=3.8\cdot10^{-5}$,
$\xi=0.02$, $\Ta\approx1.4\times10^{10}$, $\tilde{\nu}=2.9\times10^{-5}$,
and $\Pra=2.5$, resulting in $\Rey=36$, $\Co=7.6$ and $\Rat\approx3\cdot10^6$.
The discussion of the results refers to this run unless stated
otherwise.
We also consider two other runs with $\Co=4.7$ and $\Rey=39$ (Run~B3m),
as well as $\Co=14.8$ and $\Rey=31$ (Run~B5m). The former is continued from
Run~B3 of \cite{KMB11} whereas the latter is run from the initial
conditions stated above.
Our seed magnetic field has an amplitude of $\approx10^{-4}\Beq$.
As a starting point, we use $\Pm=1$ and a resolution of
$128\times256\times128$ mesh points, but also study the magnetic
Prandtl number dependence by continuing Run~B4m with values
$\Pm$=(0.25,0.5).
In Run~B4m the magnetic field grows exponentially over roughly 1500
convective
turnover times before reaching the saturated stage in which the total
rms magnetic field is $\brms=0.72\Beq$.

\begin{figure}[t]
\centering
\includegraphics[width=0.49\columnwidth]{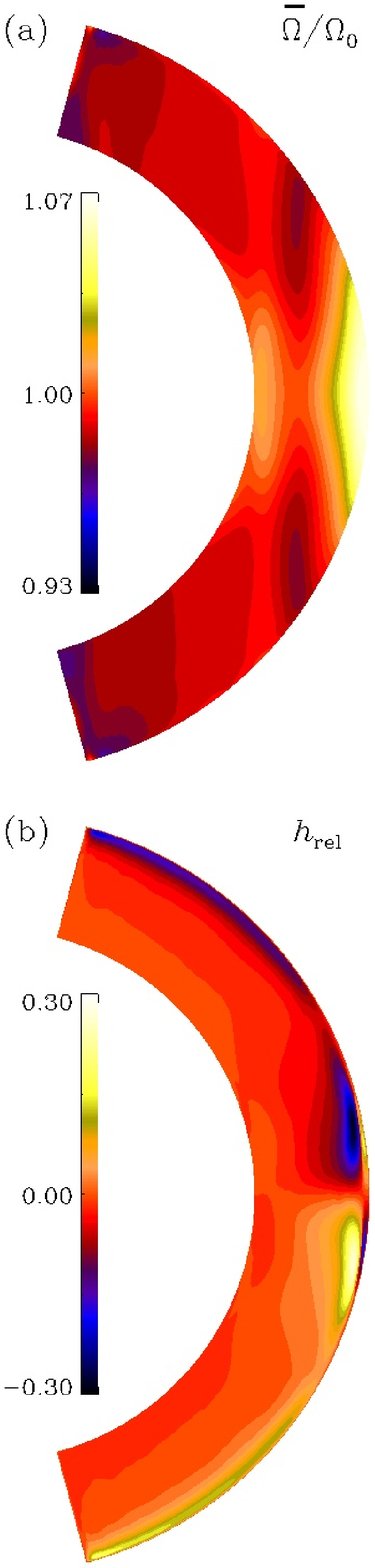}\includegraphics[width=0.489\columnwidth]{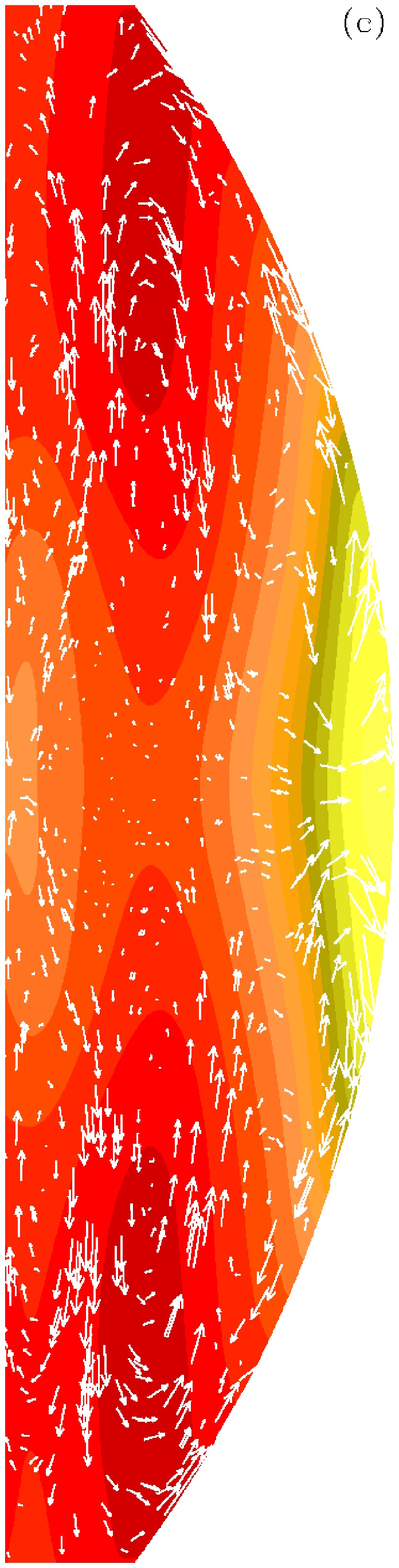}
\caption{(a) Normalized time-averaged mean rotation profile
  $\mean{\Omega}/\Omega_0=\mean{U}_\phi/(\Omega_0 r \sin \theta)+1$.
  (b) Relative kinetic helicity density $h_{\rm rel}$. (c) Rotation
  profile (color contours) and meridional circulation $\mean{\bm
    U}_m=(\mean{U}_r,\mean{U}_\theta,0)$ (arrows) near the 
  equator. From
  Run~B4m.}\label{fig:pOm}
\end{figure}

\begin{figure}[t]
\centering
\includegraphics[width=\columnwidth]{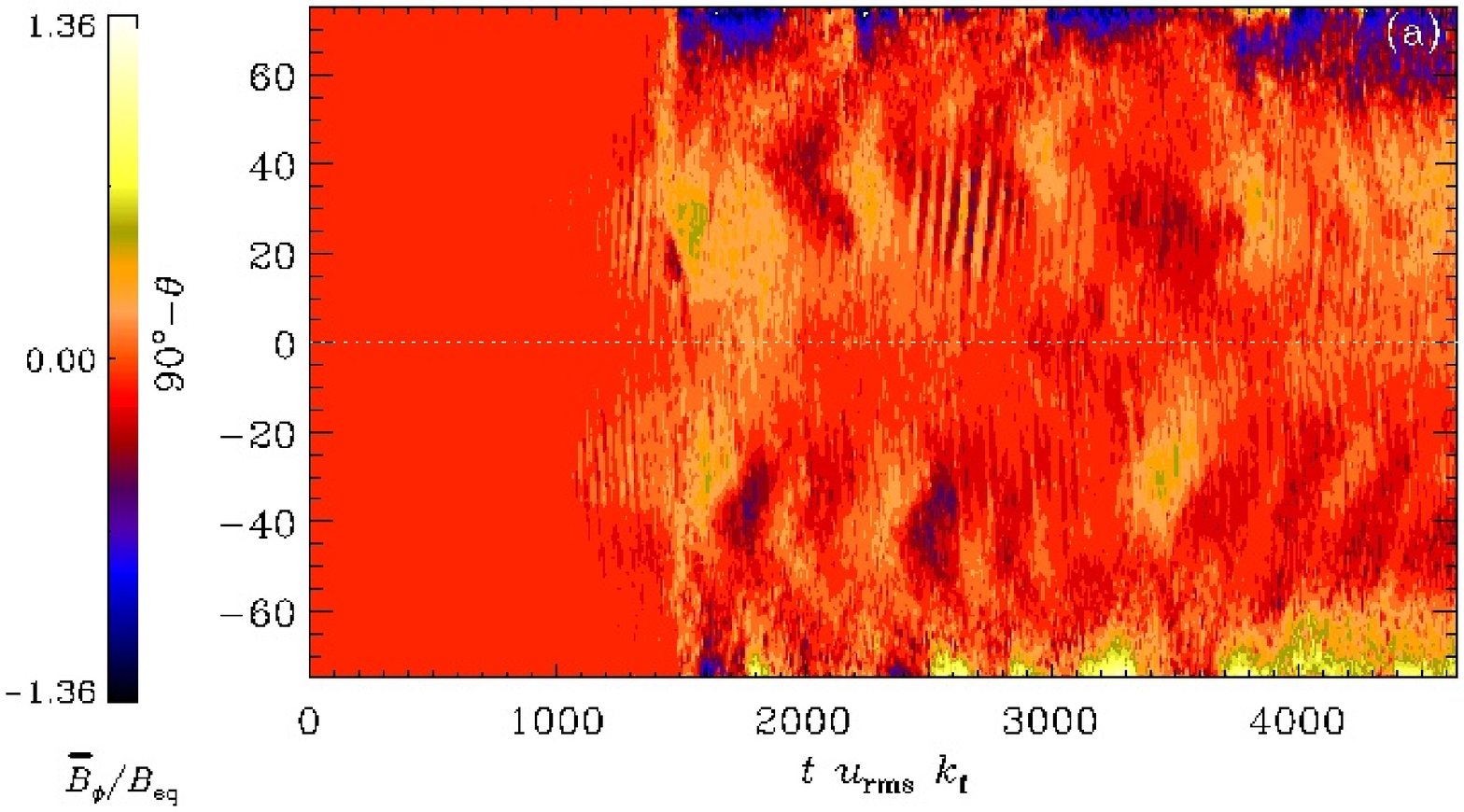}
\includegraphics[width=\columnwidth]{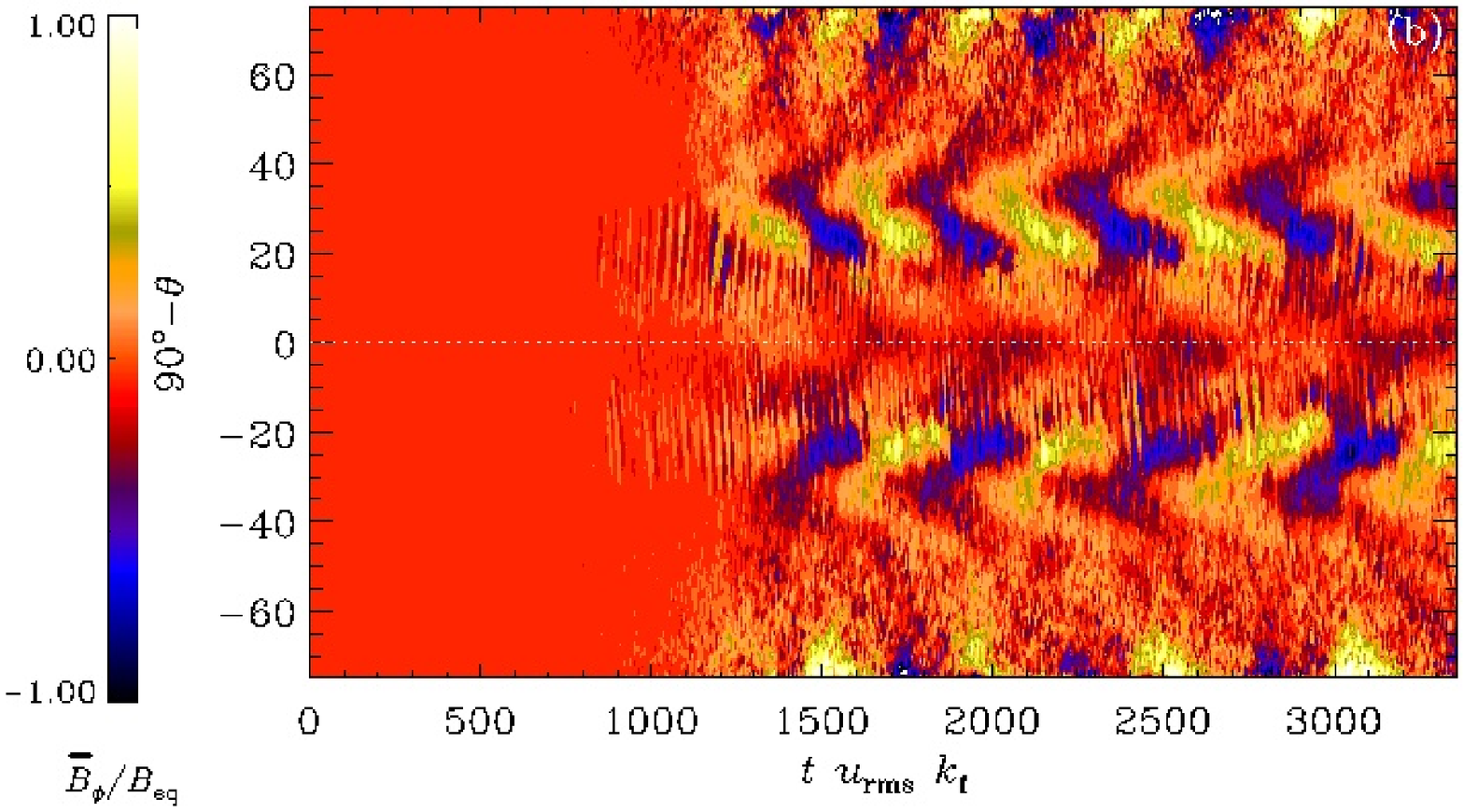}
\includegraphics[width=\columnwidth]{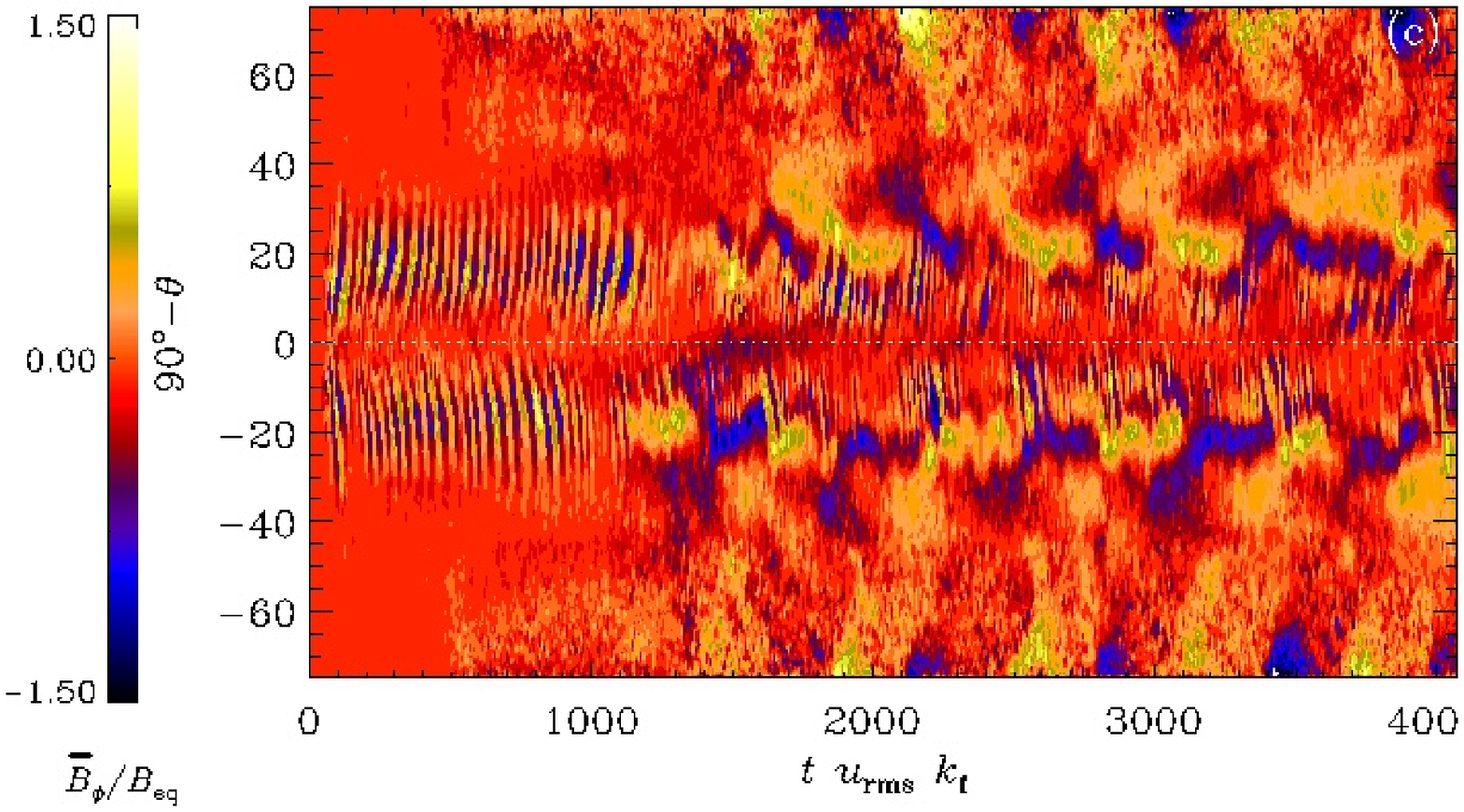}
\caption{
  $\mean{B}_\phi$ near the surface of the star at 
  $r=0.98R$ as a function of latitude $90\degr-\theta$
  for $\Co=4.7$ (top, Run~B3m), 7.6 (middle, B4m), and 14.8 (bottom, B5m).
  The white dotted line denotes the equator $90\degr-\theta=0$.
}\label{fig:butterfly}
\end{figure}

\begin{figure}[t]
\centering
\includegraphics[width=\columnwidth]{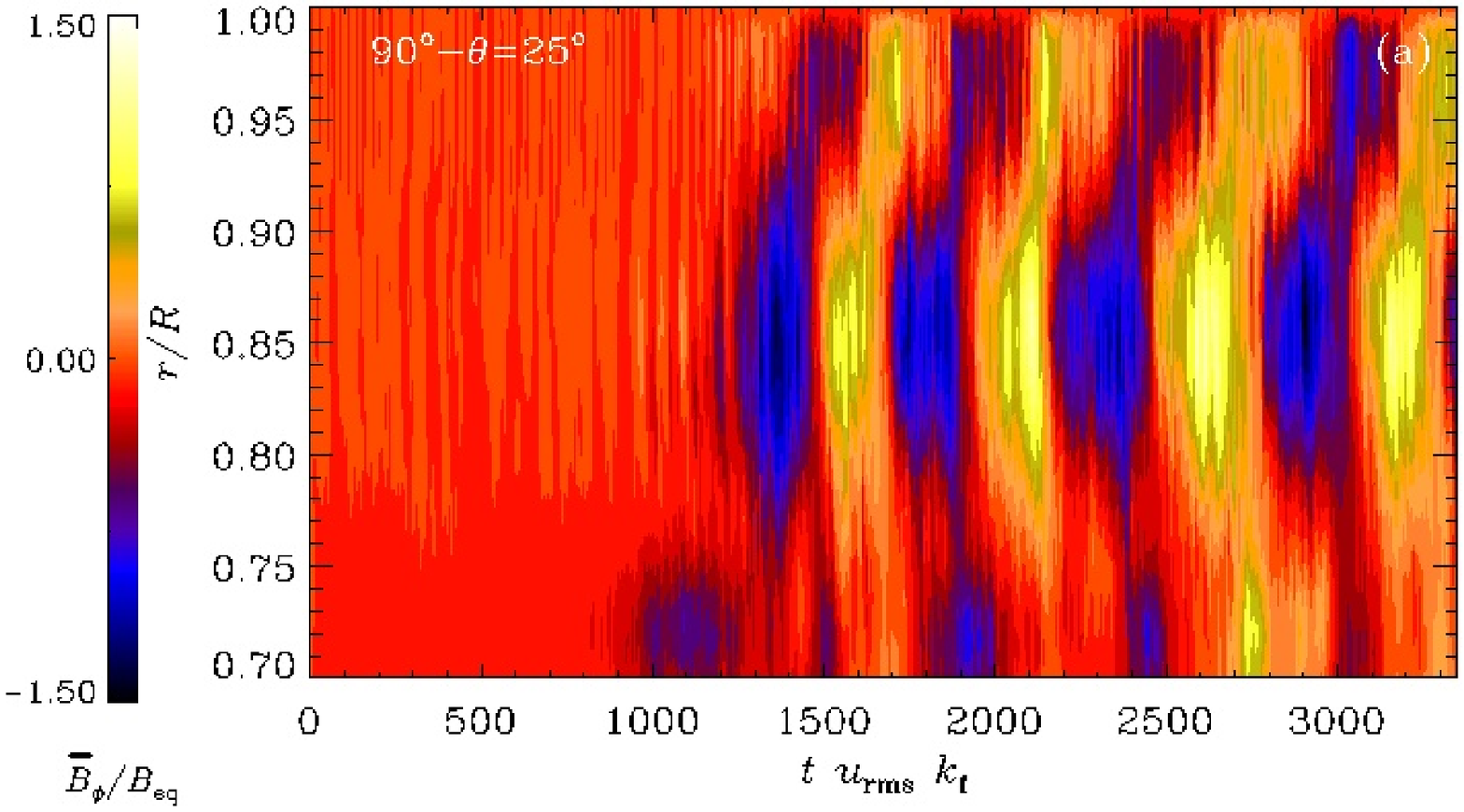}
\includegraphics[width=\columnwidth]{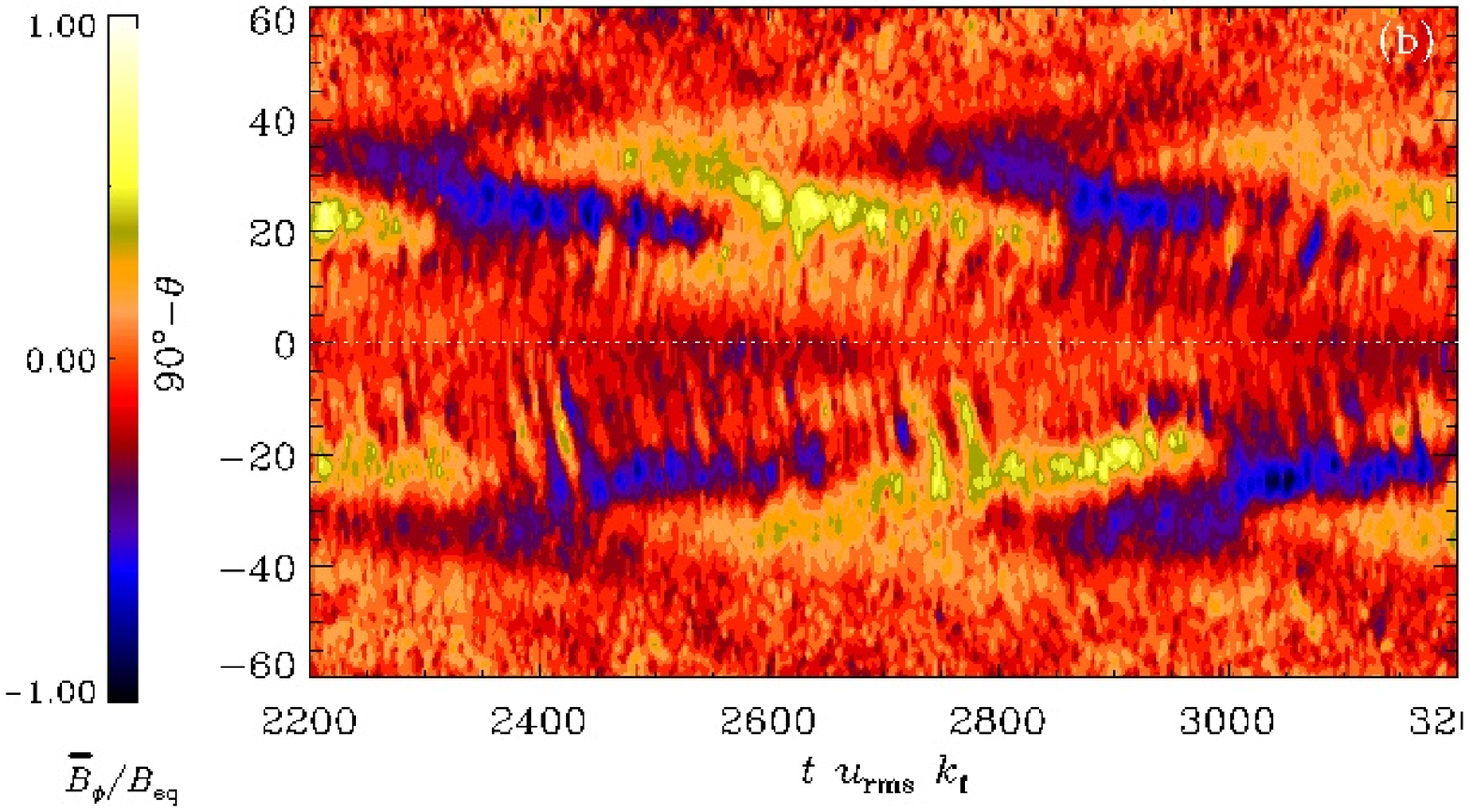}
\includegraphics[width=\columnwidth]{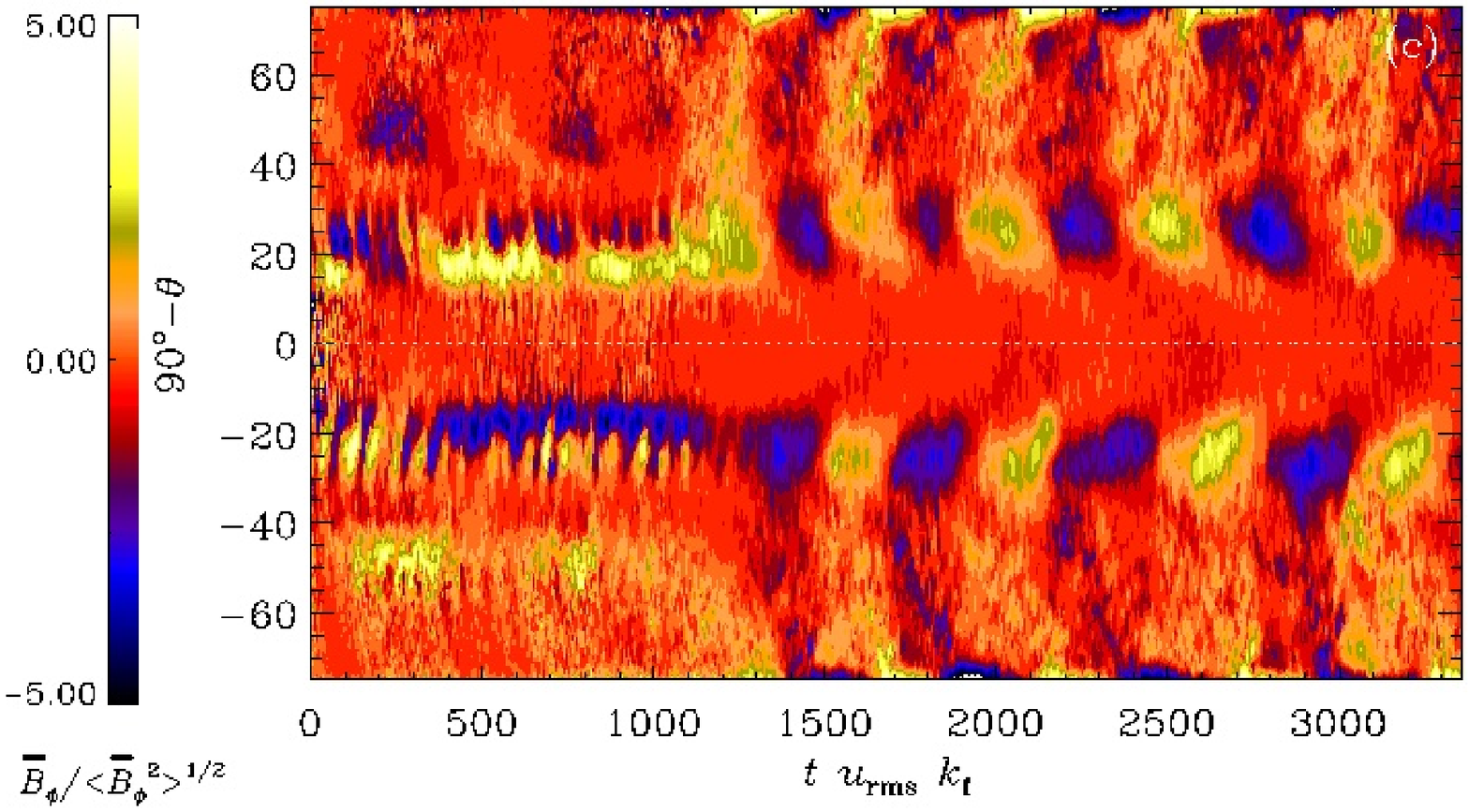}
\caption{(a) $\mean{B}_\phi(r,t)$ in units of 
  the local equipartition field strength at $25\degr$ latitude
  for Run~B4m shown in Figure~\ref{fig:butterfly}(b).
  (b) Blow-up of Figure~\ref{fig:butterfly}(b) showing
  the region $-60\degr<90\degr-\theta<60\degr$ 
  and $2200<t\urms\kef<3200$ at $r=0.98R$.
  (c) Like Figure~\ref{fig:butterfly}(b), but at $r=0.85R$
  and $\mean{B}_\phi$ is normalized by its volume averaged
  rms-value at each time to make the early time evolution visible.
}\label{fig:butterfly2}
\end{figure}

\begin{figure}[t]
\centering
\includegraphics[width=\columnwidth]{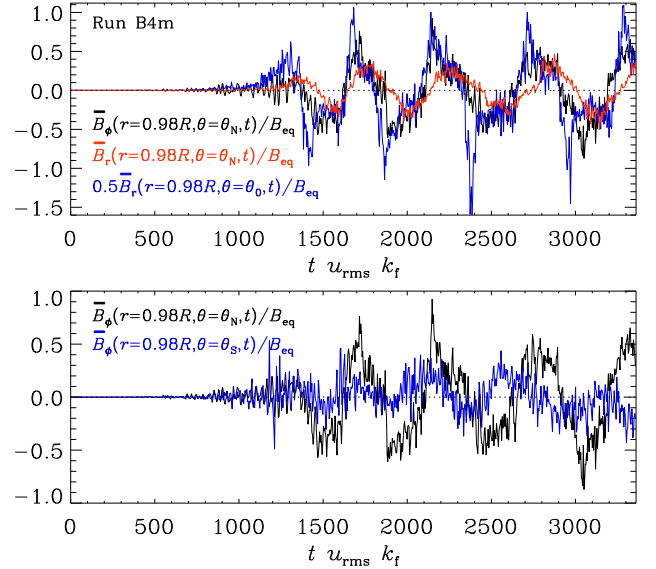}
\caption{Top panel: $\mean{B}_\phi$ (black line) and $\mean{B}_r$
  (red) at $90\degr-\theta_{\rm N}=25\degr$ latitude. The blue line 
  shows $0.5\mean{B}_r$ at $\theta_0$. Bottom panel:
  $\mean{B}_\phi$
  from $\theta_{\rm N}$ and
  $\theta_{\rm S}$ corresponding to latitudes $\pm25$ degrees,
  respectively.}
\label{fig:pline_meanB}
\end{figure}

\begin{figure*}[t]
\centering
\includegraphics[width=0.333\textwidth]{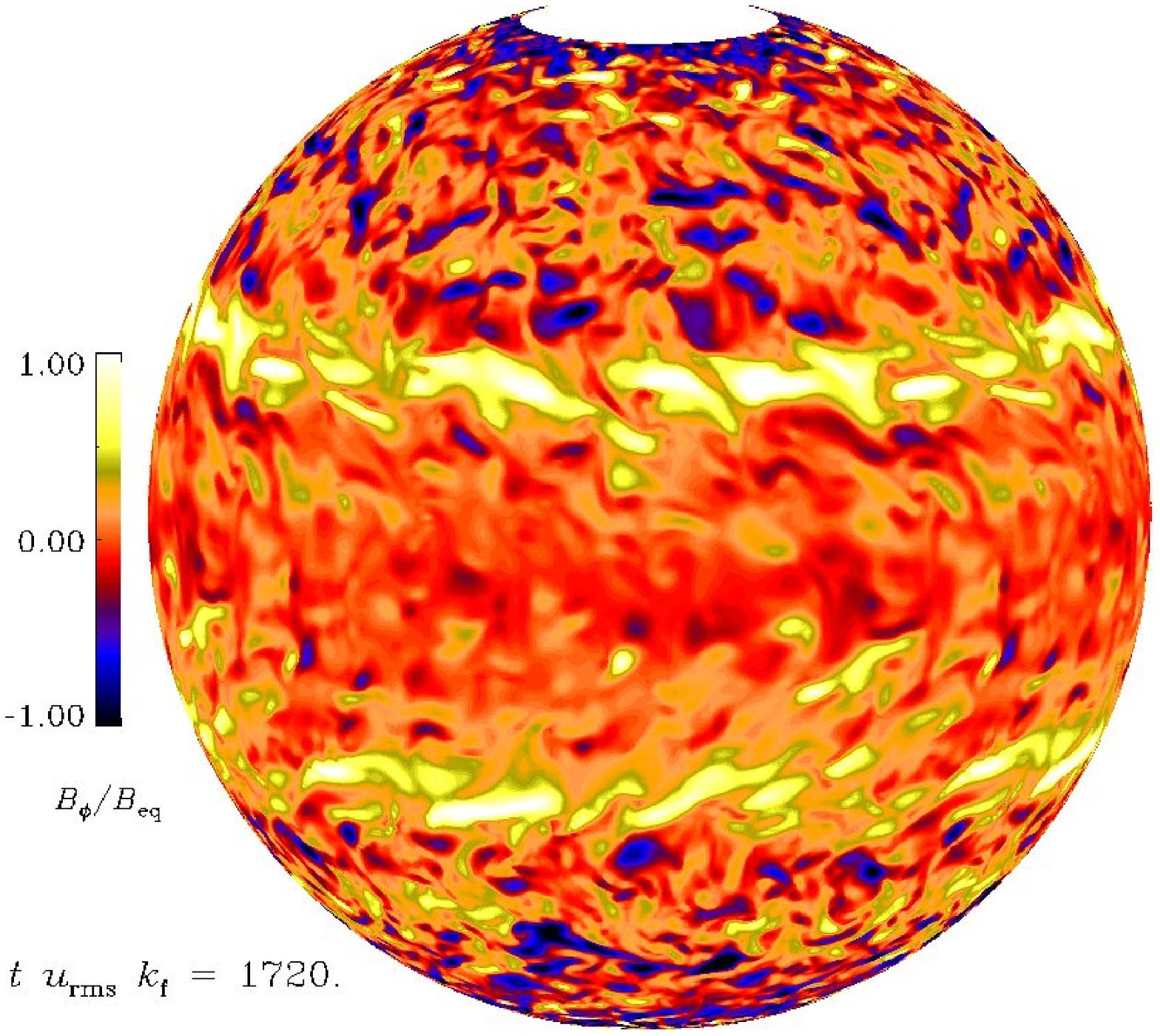}\includegraphics[width=0.333\textwidth]{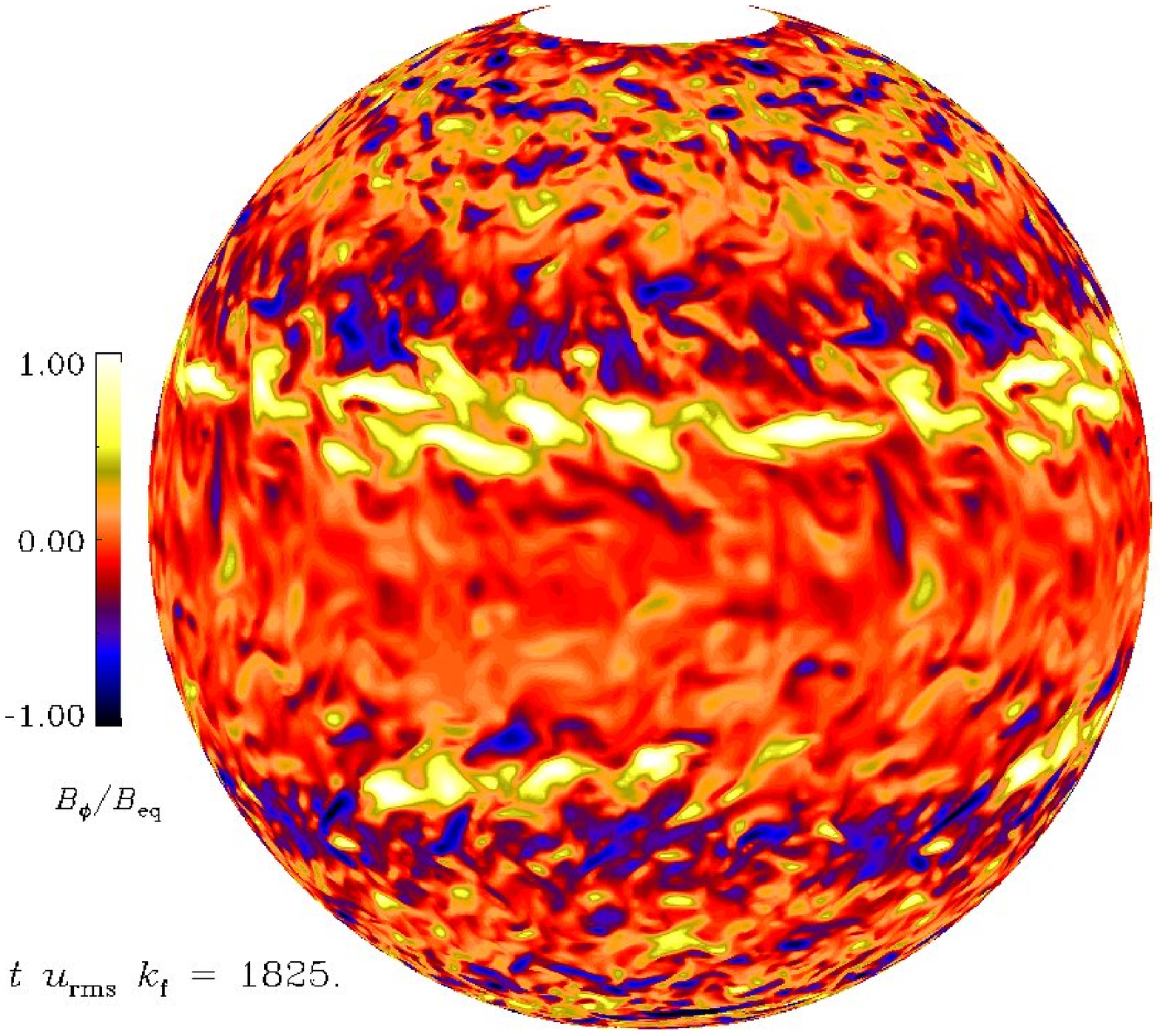}\includegraphics[width=0.333\textwidth]{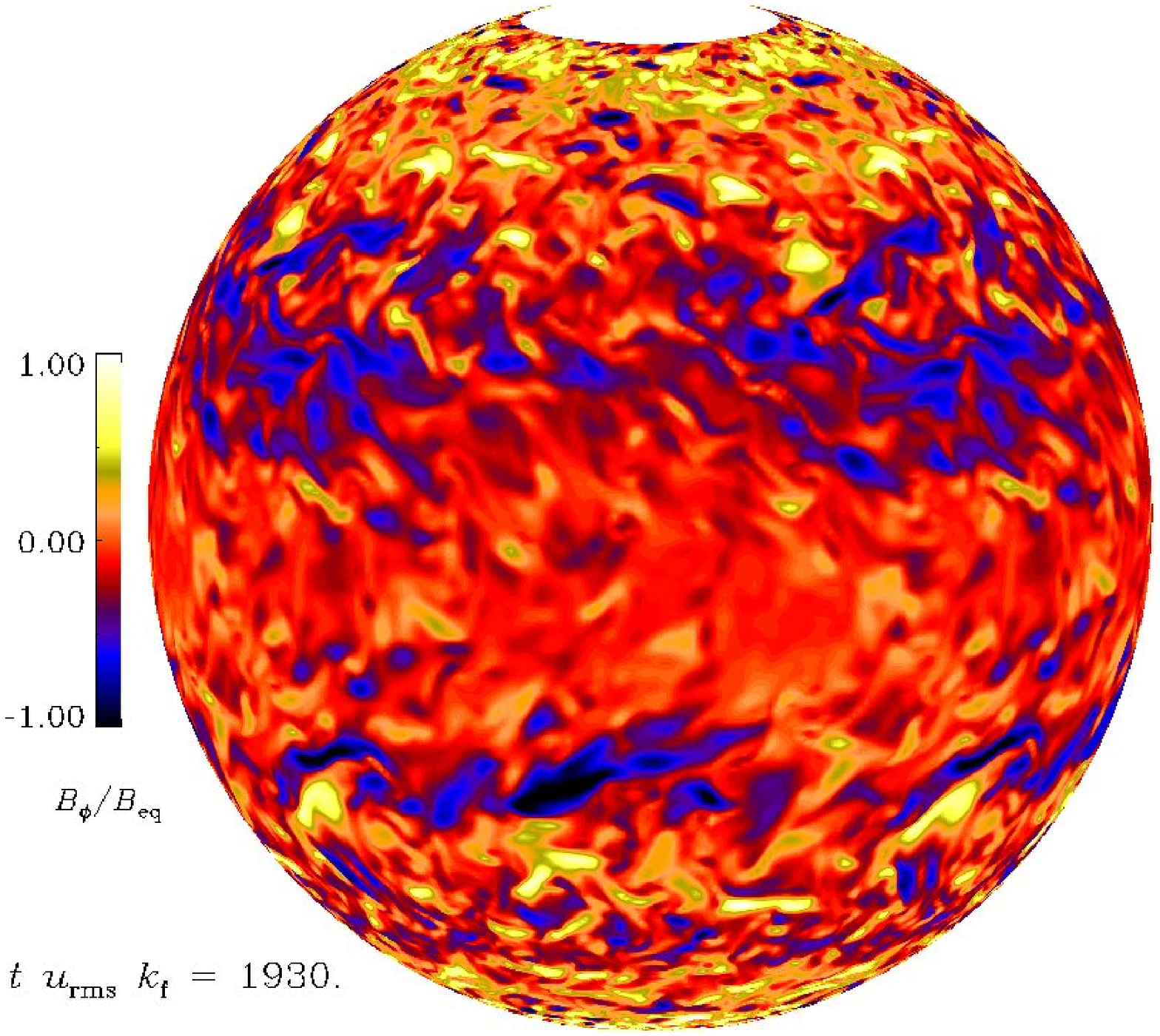}
\includegraphics[width=0.333\textwidth]{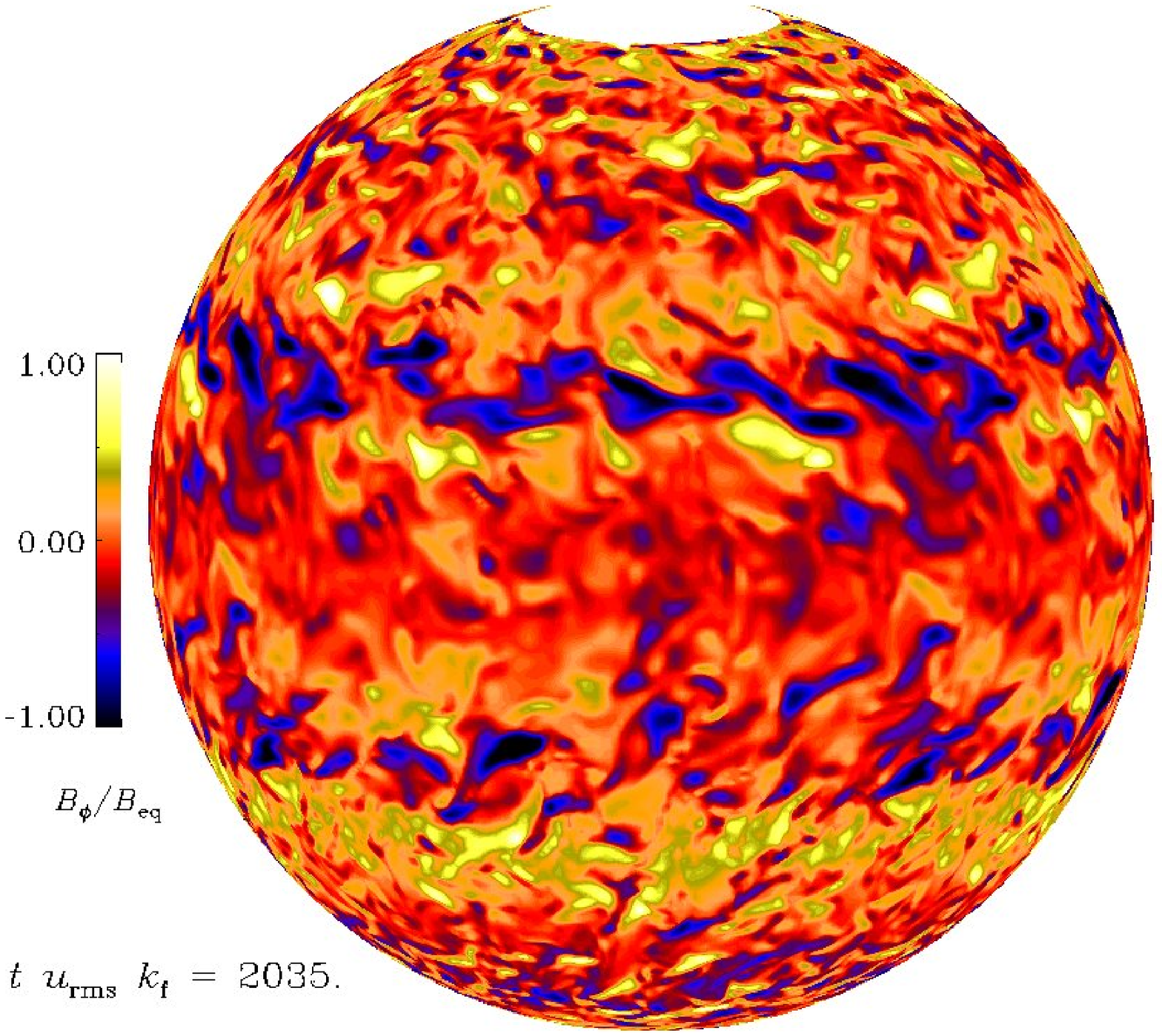}\includegraphics[width=0.333\textwidth]{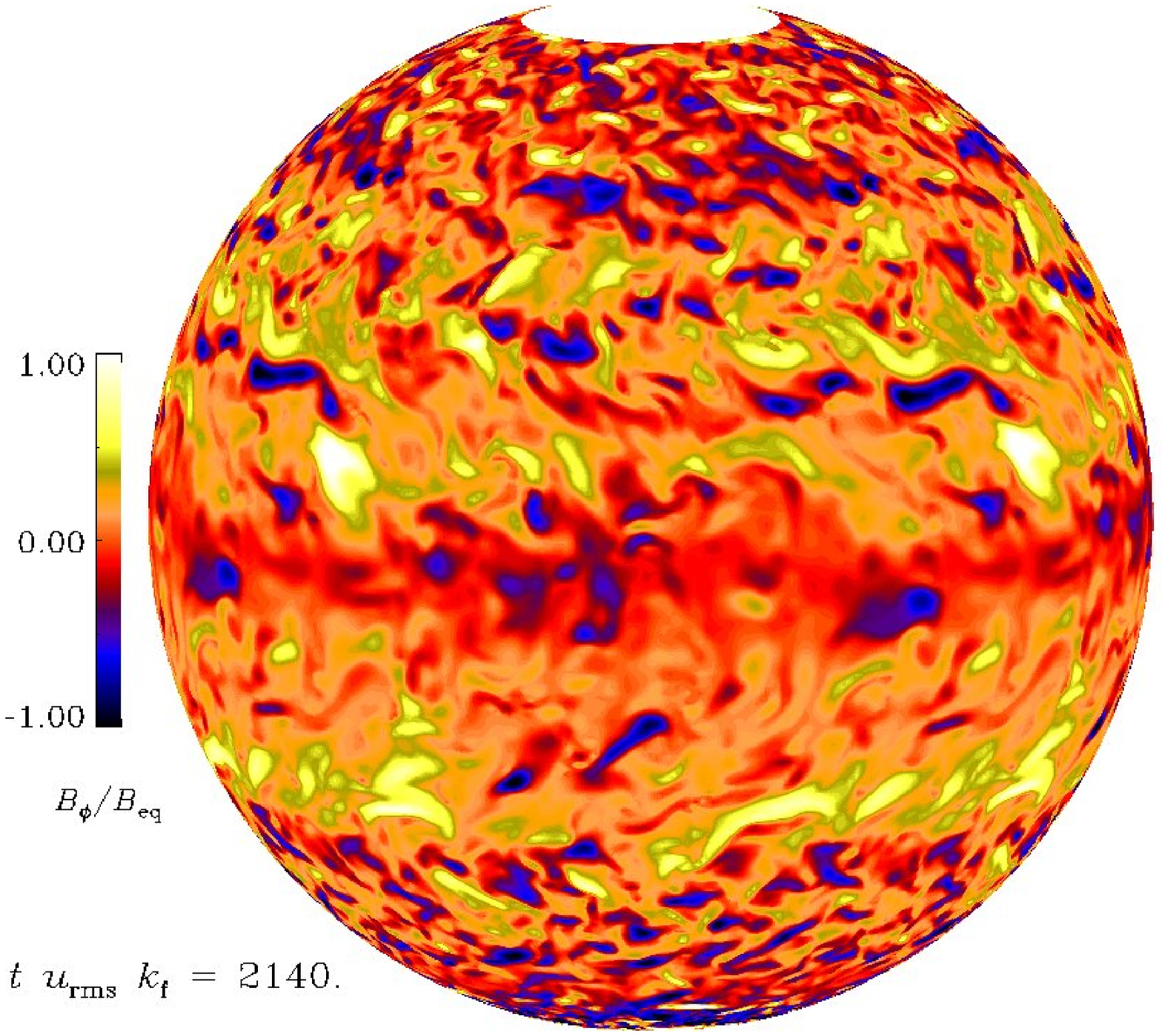}\includegraphics[width=0.333\textwidth]{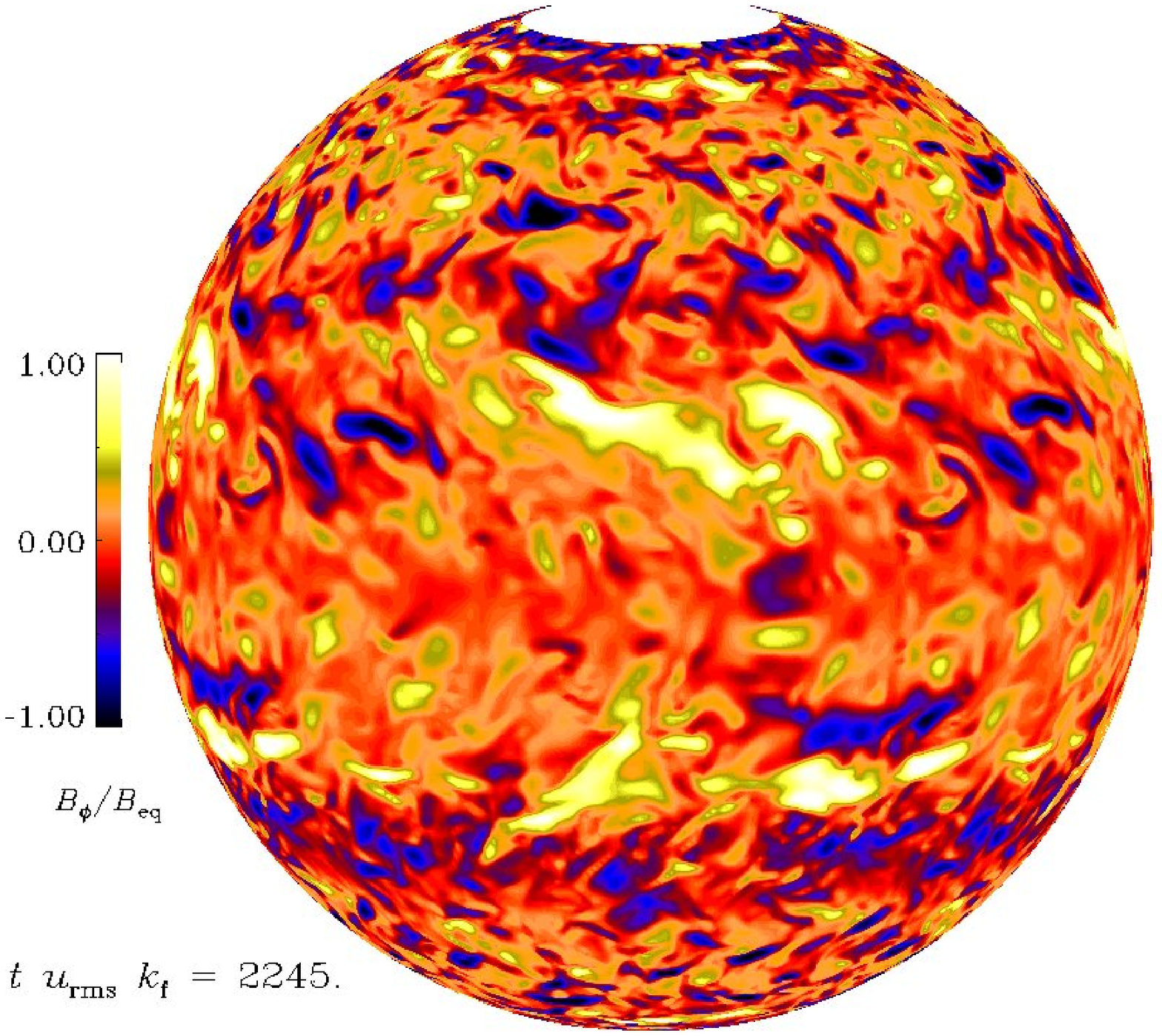}
\caption{Snapshots of the toroidal magnetic field $B_\phi$ at
  $r=0.98R$ from Run~B4m at six different times separated by $\Delta t \urms
  \kef\approx105$.}
\label{fig:Bphi}
\end{figure*}

The convection pattern near the surface shows smaller scales
at high latitudes and larger elongated structures or
`banana cells' near the equator.
Figure~\ref{fig:pOm} shows the rotation profile in the saturated
regime of the dynamo from Run~B4m. The equator rotates faster than the
high
latitudes and significant radial differential rotation is present only
near the equator. In the lower part of the convection zone, $\pd
\mean{\Omega}/\partial r$ is negative at low latitudes
(just outside the inner tangent cylinder) and
positive at high latitudes. The meridional
circulation shows several small cells outside the tangent cylinder on
both hemispheres. The latitudinal differential
rotation, measured by
$\Delta_\Omega\equiv(\Omega_{\theta=\theta_0}-\Omega_{\rm
  eq})/\Omega_{\rm eq}$, where $\Omega_{\rm eq}=\Omega(\theta=\pi/2)$,
decreases from $0.08$ in the kinematic regime to
$0.07$ in the saturated state.
The rotation profiles in Runs~B3m and B5m are qualitatively similar.

We define mean quantities as averages over longitude and denote 
them by an overbar.
In Run~B4m the relative kinetic helicity density $h_{\rm rel} 
=\mean{{\bm u} \cdot \bm\omega}/\urms \omega_{\rm rms}$, with
$\bm\omega = \bm\nabla \times {\bm u}$, is negative (positive) in the
northern (southern) hemisphere; see Figure~\ref{fig:pOm}.
No pronounced sign reversal with depth is seen.
The maximum value of $h_{\rm rel}$ is around 0.3, which
allows us to determine the dynamo number describing
the strength of the $\alpha$ effect as
$C_\alpha=\alpha/\etatz k_1\approx h_{\rm rel}\kfo/k_1\approx2.7$, where
$\kfo=\omega_{\rm rms}/\urms$ is the approximate
wavenumber of the energy-carrying eddies, $k_1=\pi/\Delta r$
is the lowest radial wavenumber in the domain,
while $\alpha\approx h_{\rm rel}\urms/3$ and $\etatz=\urms/3\kfo$
are estimates for $\alpha$ effect and turbulent magnetic diffusivity.
(We note that $\chit/\etatz$ varies between 1.9 near the surface
and 0.15 within the convection zone.)
The relevant dynamo number characterizing the radial differential
rotation is $C_\Omega=\Delta\Omega/\etatz k_1^2\approx55$, where we
have used $\Delta\Omega/\Omega_0=0.06$ for the normalized radial shear
(not to be confused with $\Delta_\Omega$ defined above).
The ratio $C_\Omega/C_\alpha$ is well over 10.
Following \cite{RS72}, this suggests that we are in what is known
as the $\alpha\Omega$ regime where shear is strong enough to
favor cyclic behavior.

Time series of the averaged longitudinal component of the
magnetic field are shown in Figure~\ref{fig:butterfly} for different values
of $\Co$=(4.7, 7.6, and 14.8).
For $\Co=7.6$, two activity
belts are visible; one propagating poleward at high latitudes
and another propagating equatorward between
10 and 30 degrees latitude.
The equatorward branch at mid-latitudes is visible also for $\Co=4.7$
but there the cycle is irregular and the magnetic field at low
latitudes weaker. The dynamo mode appears to change to a
non-oscillatory one after around $3800 \, t \urms \kef$.
In the most rapidly rotating case with
$\Co=14.8$ the cyclicity and equatorward migration of the field are
clearly present but fluctuations from one cycle to the next are
again larger than in Run~B4m.
In all of the runs a poleward branch with a shorter period is visible
near the surface at low latitudes which looks similar to the solution obtained
in the nonlinear stage in \cite{KKBMT10}. This mode can be
occasionally distinguished in the saturated stage, in particular in Run~B5m,
but it remains subdominant to the mode with longer period exhibiting equatorward
migration.

The magnetic field is strongest at $r/R\approx0.85$ and seems to propagate
from there to top and bottom of the convection zone; see
Figure~\ref{fig:butterfly2}(a) which shows $\mean{B}_\phi$ as a function of $r$
and $t$ being regenerated in the bulk of the convection zone during each cycle.
As the field approaches the surface, it propagates equatorward at low latitudes;
see Figure~\ref{fig:butterfly2}(b).
This mode becomes apparent in the nonlinear phase whereas in the
kinematic stage the solution in the bulk of the convection zone does
not oscillate; see Figure~\ref{fig:butterfly2}(c) for $t\,\urms\kef<1000$.
Since $\mean{B}_\phi$ is here normalized by the instantaneous average value,
one sees the spatio-temporal structure, and that no reversals occur.
Opposite transitions
(from oscillatory to quasi-steady) have been observed in Cartesian
simulations \citep{HRB11,KMB12}.

On theoretical grounds, we would expect $|\mean{B}_\phi/\mean{B}_r|$
to be of the order of $|C_\Omega/C_\alpha|^{1/2}\approx4.5$, but the
actual ratio is only around unity; see Figure~\ref{fig:pline_meanB}.
We cannot therefore be certain that the dynamo is really in
the $\alpha\Omega$ regime, as discussed above.
Interestingly, $\mean{B}_r$ shows a greater amplitude at high latitudes while
$\mean{B}_\phi$ is stronger at lower latitudes. Furthermore,
$\mean{B}_r$ at high latitudes changes sign approximately when
$\mean{B}_\phi$ in the low-latitude activity belt changes sign.

Visualizations of the toroidal magnetic field from Run~B4m near the
surface (Figure~\ref{fig:Bphi}) show a
persistent activity belt near the equator which is changing polarity
with a period of roughly $400\tau$ where $\tau=(\urms \kef)^{-1}$ is
the convective turnover time.
We find that decreasing $\Rm$ to 18 ($\Pm=0.5$) shortens the cycle
period by roughly 20 percent to $330\tau$. At $\Rm=9$ ($\Pm=0.25$) the
magnetic field decays, but the decay mode is oscillatory with a period
of $270\tau$. Similar increase of the cycle period with $\Rm$ has been
observed in forced turbulence simulations \citep{KB09} and suggests
that the current runs are not in a regime where the molecular
diffusivities are unimportant.
The same cyclic behavior is seen throughout the depth of the 
convection zone above $r=0.75R$.
Relating the turnover time of our highest $\Rm$ model
to that of the deep layers of
the solar convection zone, i.e.\ one month, leads to a magnetic cycle
period of roughly 33 years.
The cycle might well be shorter if the relevant depth is shallower.
On the other hand, if we used $\kfo$ instead of $\kef$, our cycle period
would be 4--5 times longer.
It is also noteworthy that
$\mean{B}_\phi$ has mixed parity about the equator, except around the
time $t\urms\kef=2500$ when the field is of odd parity; see
Figure~\ref{fig:pline_meanB}.

\section{Conclusions}
\label{sec:conclusions}
We have reported solar-like magnetic cycles from simulations of turbulent
convection in spherical wedge geometry. The magnetic activity is
concentrated in two belts, a high-latitude one propagating
poleward, and a low-latitude one propagating equatorward. The strongest
magnetic fields, however, occur in the high-latitude activity branch.
Simulations with moderately slower and faster rotation show similar
behavior. These results will be discussed in more detail in forthcoming 
publications.
Relating
the convective turnover time in the simulation to that of the Sun we
obtain a cycle period of 33 years which is somewhat longer than that in the
Sun and half that obtained by \cite{Ghizaru10} from quite a different
model exhibiting similar solutions, but without equatorward migration.
One of the main differences to our earlier work \citep{KKBMT10} is
that we have omitted a stably stratified overshoot layer beneath. This
allowed us to cover almost an order of magnitude larger density
contrast within the convectively unstable layer. Furthermore,
convective energy transport now dominates over radiative diffusion and
a black body boundary condition is used for the temperature
\citep[cf.][]{KMB11}.
However, compared with the Sun, our contours of differential rotation are still
too cylindrical and also the banana-cell pattern of radial velocity
might not be realistic.
Both may be a consequence of having a large Taylor number;
even the turbulent Taylor number,
$(\nu/\nut)^2\Ta=9\,\Ta/\Rey^2\approx10^{8}$, is rather large.
Here, $\nut\approx\etatz$ has been used as an estimate of the
turbulent viscosity.
The magnetic activity in our model is distributed throughout
the convection zone, in contrast to the widely accepted flux-transport
dynamo mechanism \citep{DC99} in which a one-cell anti-clockwise
(north) meridional circulation is crucial.
In our model, meridional flows are convergent toward low latitudes
(see Figure~\ref{fig:pOm})
and may contribute to the resulting equatorward migration.
On the other hand, the negative radial differential rotation in
the near-surface shear layer \citep[as anticipated by][]{B05}, which
is here absent, is not the explanation for the resulting equatorward
migration. Clarifying this is an important goal for future work.

\acknowledgements
The authors acknowledge the anonymous referee for constructive
comments on the manuscript.
The simulations were performed using the supercomputers hosted by CSC
-- IT Center for Science Ltd.\ in Espoo, Finland, who are administered
by the Finnish Ministry of Education. Financial support from the
Academy of Finland grants No.\ 136189, 140970 (PJK) and 218159, 141017
(MJM), as well as the Swedish Research Council grant 621-2007-4064,
and the European Research Council under the AstroDyn Research Project
227952 are acknowledged.
The authors thank NORDITA for hospitality during their visits.

\bibliography{paper}

\end{document}